\documentclass{iopart}
\usepackage{epsfig}  
\begin{document}

\font\schwell=schwell at 12pt
\renewcommand{\l}{\schwell L\/}

\title{$\mbox{\boldmath{$\it\/J\//\psi$}}$ Propagation in Hadronic
Matter}

\author{Kevin Haglin\dag\footnote[3]{email: haglin@stcloudstate.edu} 
and Charles Gale\ddag\footnote[4]{email: gale@hep.physics.mcgill.ca}
}

\address{\dag\ Department of Physics, Astronomy and Engineering Science,
Saint Cloud State University, 720 Fourth Avenue South,
St. Cloud, MN 56301, USA}

\address{\ddag\ Department of Physics, McGill University, 
3600 University Street, Montr\'eal, QC, H3A 2T8, Canada}

\begin{abstract}
We study {\it J\/}/$\psi$ propagation in hot hadronic matter using
a four-flavor chiral Lagrangian to model the dynamics and using
QCD sum rules to model the finite size effects manifested
in vertex interactions through form factors.  Charmonium breakup due to 
scattering with light mesons is the primary impediment to continued 
propagation.  Breakup rates introduce nontrivial temperature and momentum
dependence into the {\it J\/}/$\psi$ spectral function.  
\end{abstract}




\section{Introduction}
Suppression of the {\it\/J\/}/$\psi$ signal in heavy ion collisions
is still considered one of the promising probes
of our attempts to push back deconfinement and open up
a four volume of quark-gluon plasma (QGP) from the vacuum wherein 
quarks and gluons have their confinement scale extended to
macroscopic lengths.  The greatest four
volume that one can hope for in the laboratory at present is about 
10$^{4}$ fm$^{4}$ since stable nuclei are of limited size 
($\sim$ 10 fm) and system longevity is expected to be of the 
order 10 fm. 
Any yet, dynamical scales for strong interaction physics are at
the femtometer and subfemtometer level, and so it is quite reasonable to 
approximate the system as being relatively large, relatively
long-lived, and near thermal equilibrium.  Then, kinetic theory of 
asymptotic states and vacuum cross sections is a useful tool to 
establish the baseline rates for various dynamical effects.

The mechanism for suppression of $J/\psi$ suggested by Matsui 
and Satz\cite{ms86} originates from color screening in QCD plasma 
which is clearly absent in hot hadronic matter.  Screening would 
effectively ``partonize'' the bound state mesons (the QCD analog of 
ionization in atomic physics) into open subsystems with colour (quarks, 
antiquarks and glue).  For charm, this possibility brings 
forward a probe of open ($D$ mesons) versus hidden ($J/\psi$). 
Open charm decays mostly semileptonically, and will therefore be absent in
an invariant mass window of say, muon pairs near the $J/\psi$ mass.
Therefore, a partial or total absence of a strong $J/\psi$ peak could
indicate the presence of QGP.

In order for the probe to be useful, hidden
charm must be relatively unreactive to hadrons to survive the
hadronic stage of the fireball.
For instance, 
we do not expect suppression of the $\phi$ to be a QGP signal because
the $\phi$ interacts with light mesons such that its mean
free path is of the order 2--5 fm\cite{kh95,vk02}.  We ask, could the same 
effects spoil charm--$J/\psi$ that most likely spoil strangeness--$\phi$
from utility?
Our purpose here is to assess within an effective field theory the
interaction cross sections of $J/\psi$ with light mesons (pions,
rho mesons, etc.) and to further estimate the breakup rates.
The hope is that such activity might shed some light on the extent
to which suppression of the $J/\psi$ as a plasma indicator
is emulated by purely hadronic effects.

\section{Chiral Lagrangian}

Quantum chromodynamics at low energies is particularly challenging.
Lattice technology, for example, has not yet advanced to the stage 
where one can do dynamical studies at finite temperature.  We instead
resort to very successful methods of effective Lagrangian
descriptions for the hadronic degrees of freedom.  While the symmetry
arguments typically introduced for two light quark flavours are not as
clearly applicable when generalizing to three and four flavours, we
indicate, as is usually done, that symmetry breaking effects
are relegated entirely to physical mass eigenstates, leaving the 
interaction completely symmetric from the beginning.  The 
guiding principles are therefore chiral symmetry, unitarity, and 
gauge invariance.

We start with a nonlinear sigma model
\begin{eqnarray}
{\cal\/L\/} & = & 
-{F_{\pi}^{\,2}\over\/8}\,\partial_{\mu}U\/\partial^{\,\mu}\/U^{\dag},
\end{eqnarray}
where $U = \exp\,(2\,i\phi/\/F_{\pi})$, and $\phi$ is
the pseudoscalar meson matrix appropriately
generalized to include charm, thereby becoming 4$\times$4,
and $F_{\pi}$ = 135 MeV is the pion decay constant.
A chiral covariant derivative $\partial_{\mu}U\to{D}_{\mu}U$
introduces vector and axial vector fields and their respective coupling to 
pseudoscalars.  Kinetic energy terms for spin 1 fields and generalized
mass terms of the left- and right-handed fields $A_{\mu}^{L}$ and
$A_{\mu}^{R}$ are introduced.  Then, the axial 
vector fields are gauged away leaving to lowest order a set of interactions for
light plus heavy pseudoscalar and vector mesons.  The interactions
are compactly written as\cite{khcg01}
\begin{eqnarray}
{\cal L\/}_{\rm\,int} & = & i\,g\,{\rm\/Tr\,}\left(\rho_{\mu}\/\left[\partial^{\mu}\phi,\,\phi\right]\right) - {g^{\,2}\over\/2\/}{\rm\/Tr\/}\left(
\left[\phi,\,\rho^{\mu}\,\right]^{\,2\/}\right)
+ i\,g\,{\rm\/Tr\/}\left(\partial_{\mu}\rho_{\nu}\left[\rho^{\mu},\,\rho^{\nu}
\,\right]\right)
\nonumber\\
& \ & + {g^{\,2}\over\/4\/}
{\rm\/Tr\/}\left(\left[\rho^{\mu},\,\rho^{\nu}\right]^{\,2\/}\right)\,,
\end{eqnarray}
where $\rho^{\mu}$ is also generalized to encompass charm and 
is therefore a 4$\times$4 matrix of vector mesons.
If the symmetry were perfectly realized in nature, the chiral
coupling constant $g$ would alone be sufficient to describe the
interaction strengths universally.  We relax that requirement
here and appeal to experimental results and model calculations
via QCD sum rules for fitting the coupling constants at each 
three-point vertex.  Gauge invariance then restricts the
four-point functions.

A recent measurement of $\Gamma(D^{*\,\pm}\to\,D\,\pi)$ =
96 $\pm$4 $\pm$ 22 keV\cite{cleo} fixes the value of the coupling constant
$g_{D^{*}D\pi}$ = 6.39.  Coupling of the $J/\psi$ to $D\,D$ and to
$D^{*}\,D^{*}$
vertices is also required as input to calculate the
pion-included breakup cross section for $J/\psi$.  Using strict 
vector dominance, which has been done several times in the 
past\cite{mm98,kh00,lk00,oh01},
the coupling constants would be $g_{J/\psi\,DD\/}$ =
$g_{J/\psi\,D^{*}D^{*}\/}$ = 7.7.   In the next section we
indicate an alternative approach of extracting not only 
the coupling strengths but also momentum dependence at
the vertices from QCD sum rules.  The
$\rho$-induced inelastic reactions require further information
via coupling constants.  Vector meson dominance and chiral
symmetry would lead to $g_{\rho\,D\,D\/}$ =
$g_{\rho\,D^{*}\,D^{*}\/}$ = 5.6\cite{mm98}.  We alternatively
look to QCD sum rules for information on the coupling strengths and 
momentum dependences, described below.

\section{Form Factors}

Hadrons are clearly composites of the underlying quarks whose
(effective) fields describe pointlike physics only when all the
interacting particles are on mass-shell.  As soon as at least one of 
the particles in a vertex is pushed kinematically off-shell, the finite
size effects become important.  This is of course not true
for gauge field theories ({\it\/e.g.\/} QED, QCD) whose
degrees of freedom are structureless.  One way to account for this is to 
use the effective Lagrangian to construct higher order processes (vertex 
corrections, loop corrections to propagators, etc.) and to formulate
strategies for renormalization.  This has not been done up to now.  However, 
the same physics has been modelled using QCD sum rules to extract the 
momentum dependence of the vertex coupling constants which then
``run'' with the relevant squared four momenta contacting the 
vertex\cite{na00}.
The basic scheme is to use on the one hand, Wilson's operator product 
expansion to write the vertex, and on the other, phenomenolgy
in the form of Borel transformations to write the same function as linear
combinations of relevant four momenta.   Running coupling constants
have been obtained as follows\cite{na02,ma02,br02}

\begin{eqnarray}
g_{DD\/J/\psi}(t) & = & g_{DD\/J/\psi}\,
e^{-\,\left[\left(t-16.2\right)^{\,2}/228\right]}
\ \equiv\ \,h_{1}(t)
\nonumber\\
g_{D^{*}\/D\/\pi}(t) & = & g_{D^{*}\/D\/\pi}\,\left({\Lambda_{D}^{2}-m_{D}^{2}
\over\Lambda_{D}^{2}-t}\right)
\hskip0.55cm \equiv \ \,h_{2}(t)
\nonumber\\
g_{D\/D\/\rho}(t) & = & g_{D\/D\/\rho}\,\left({\Lambda_{D}^{2}-m_{D}^{2}
\over\Lambda_{D}^{2}-t}\right)
\hskip0.75cm \equiv \ \,h_{3}(t)
\end{eqnarray}
where $t$ is the squared four momentum of the off-shell $D$ meson,
$\Lambda_{D}$ = 3.5 GeV, $g_{D\/D\/\rho}$ = 4.4, 
$g_{D^{*}\/D\/\pi}$ = 6.39,
and where $g_{D\/D\/J\psi}$ = 
16.4.  These values are different by tens of percents from the chiral
symmetry assumption and strict vector meson dominance.

\begin{figure}[!t]
\begin{center}
\epsfig{file=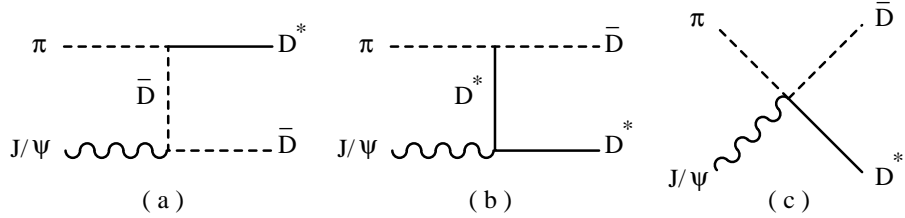,width=4.5cm}
\hspace{0.5cm}
\end{center}
\caption{Lowest order Feynman diagrams for the
process $J/\psi+\pi\to\/D^{*}\bar{D}$.  As is typical in these
theories, there is a {\it\/t\/}-channel (a), a {\it\/u\/}-channel
(b) and a four-point graph (c).}
\label{feynf}
\end{figure}

We next illustrate our gauge invariant method of implementing form factors
for the channel $J/\psi + \pi \to D^{*} + \bar{D}\/$.  Notice from
Fig.~\ref{feynf} that there
are three Feynman graphs contributing to the scattering
amplitude: a {\it\/t\/}-channel, {\it\/u\/}-channel, and a four-point
graph.
The full amplitude becomes
\begin{eqnarray}
{\cal M\/} & = & {\cal\/M\/}_{t}^{0}\,h_{1}(t\/)\,h_{2}(t\/)\,+\,
                 {\cal\/M\/}_{u}^{0}\,h_{1}(u\/)\,h_{2}(u\/)\,+\,
 {\cal\/M\/}_{\rm\,4\,pt}^{0}\left(-g^{\mu\nu}\to\/X^{\mu\nu}\right)
\end{eqnarray}
where the superscript ``0'' indicates amplitudes
without form factors (without running couplings) modulo
the coupling constants, and
where the four-point amplitude is evaluated
only after replacing ``$-g^{\,\mu\nu}$'' with $X^{\mu\nu}$,
where
\begin{eqnarray}
X^{\mu\nu} & = & A\,g^{\mu\nu} 
+ B\,\left(p_{D}^{\mu}p_{\pi}^{\nu}+ p_{\pi}^{\mu}p_{D}^{\nu}\right)
+ C\,\left(p_{D^{*}}^{\mu}p_{\pi}^{\nu}+ p_{\pi}^{\mu}p_{D^{*}}^{\nu}\right)
\nonumber\\
& \ &
+ D\,\left(p_{\pi}^{\mu}p_{\pi}^{\nu}+ p_{D}^{\mu}p_{D}^{\nu}\right)
+ E\,\left(p_{\pi}^{\mu}p_{\pi}^{\nu}+ p_{D^{*}}^{\mu}p_{D^{*}}^{\nu}\right).
\end{eqnarray}
Gauge invariance restricts the expansion coefficients $A,\ldots\,E\/$, albeit
not uniquely.  A Lorentz and gauge-invariant solution is the following
\begin{eqnarray}
A & = & -\,h_{1}(t)\,h_{2}(t)
\nonumber\\
B & = & D \ \ = \ \ {h_{1}(t)\,h_{2}(t)\,-\, h_{1}(u)\,h_{2}(u)\over
                     p_{J/\psi}\cdot\,p_{\pi}+ p_{J/\psi}\cdot\,p_{D}}
\nonumber\\
C & = & E \ \ = \ \ 0.
\end{eqnarray}

This strategy accommodates extended hadrons and is applied to all
processes, including $\rho$-induced channels.
Before coming to the results, we reiterate the
features of our form-factor implementation 
and underscore the importance of using 
\begin{itemize}
\item[$\bullet$]form factors which are covariant,
\item[$\bullet$]unique form factors at each vertex which are appropriate for 
the specific kinematics introduced at that vertex (rather than some 
average momentum transfers over all graphs), and
\item[$\bullet$]an approach where gauge invariance is strictly respected.  
\end{itemize}
Giving up on Lorentz invariance, gauge
invariance, or failing to insist on the proper kinematics for each 
form factor should be considered less reliable and therefore unacceptable.

\section{Results}

\begin{figure}[!t]
\begin{center}
\epsfig{file=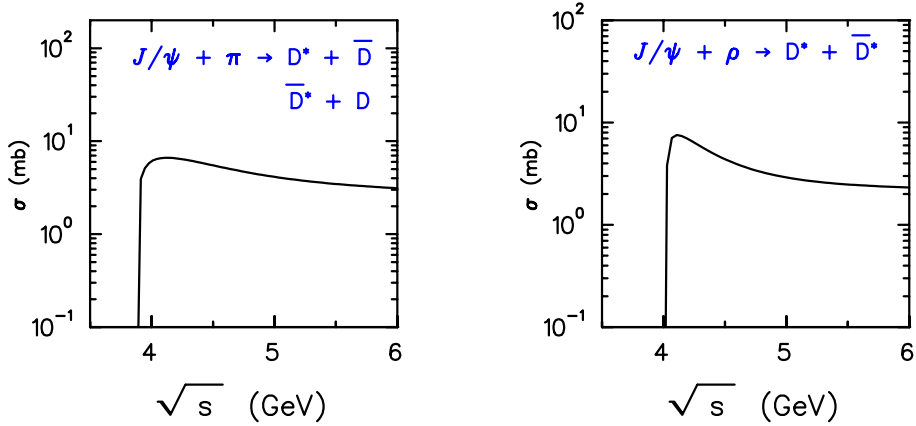,width=4.5cm}
\hspace{0.05cm}
\end{center}
\caption{$\sigma(J/\psi+\pi\to\/D^{*}\bar{D}+
\/\bar{D}^{*}\,{D})$
and $\sigma(J/\psi+\rho\to\/D^{*}\bar{D}^{*})$.}
\label{xsections}
\end{figure}

Cross sections for $J/\psi$ dissociation are inaccessible experimentally.
It is therefore crucial to have reliable estimates for the magnitude
and energy dependence of such quantities.  Only then can we
address the earlier question regarding the extent to which the $J/\psi$ 
is unreactive to light hadrons.  We report in 
Fig.~\ref{xsections} the results of some of the 
pion and rho meson induced breakup cross sections as functions of energy. 
The near-threshold and energy
dependences are nontrivial and the magnitudes of the
cross sections are several millibarns.  We note that quark-exchange models
have recently given cross sections of the order 1 mb\cite{wsb02}. 
We further note that while the magnitude of the cross sections
here are similar to other recent meson-exchange models, the energy dependence
is quite different owing to different form factors and different
implementation\cite{lk02,nnck02,du02,oslw02}.

Reaction rates within a fireball $\Gamma\,(\omega,\,\vec{\,p}\,)$ 
are also important indicators of spatial and temporal 
scales\cite{kh95,we92,sg99}.   Here we use the formalism from
Ref.\cite{khcg01} to write the  $J/\psi$ spectral
function as
\begin{eqnarray}
A_{J/\psi} (\omega, \vec{\,p}\,)\ & = & \ - 2\,{\rm Im} D_{J/\psi} (\omega,
\vec{\,p}\,)\ ,
\end{eqnarray}
where $D_{J/\psi}$ is the scalar part of the $J/\psi$ propagator.
If we neglect the difference between longitudinal and transverse
polarizations of the $J/\psi$ in the finite temperature medium
\cite{gk91}, then we have
\begin{eqnarray}
D_{J/\psi} (\omega, \vec{\,p}\,)\ & = & \ \frac{1}
{p^{\,2} - m_{J/\psi}^2 - F (\omega, \vec{\,p}\,)}\ ,
\end{eqnarray}
where $p^{\,\mu} =  (\omega, \vec{\,p}\,)$, and $F$ is the scalar 
imaginary self-energy.  Then, using 
\begin{eqnarray}
\Gamma_{J/\psi}=- 1 / m_{J/\psi}~{\rm Im}~F({p^{\,2\/}}=m_{J/\psi}^2)\,,
\end{eqnarray}
we write
\begin{eqnarray}
A_{J/\psi} (\omega, \vec{\,p}\,)\ = \ \frac{2\, m_{J/\psi}\, 
\Gamma_{J/\psi}}{(p^{\,2} - m_{J/\psi}^2)^{\,2} + m_{J/\psi}^{\,2}
\Gamma_{J/\psi}^{\,2} }\ ,
\end{eqnarray}
where $\Gamma_{J/\psi}$ contains the vacuum width and the 
contributions from inelastic scattering with pions and rhos.

\begin{figure}[!t]
\begin{center}
\epsfig{file=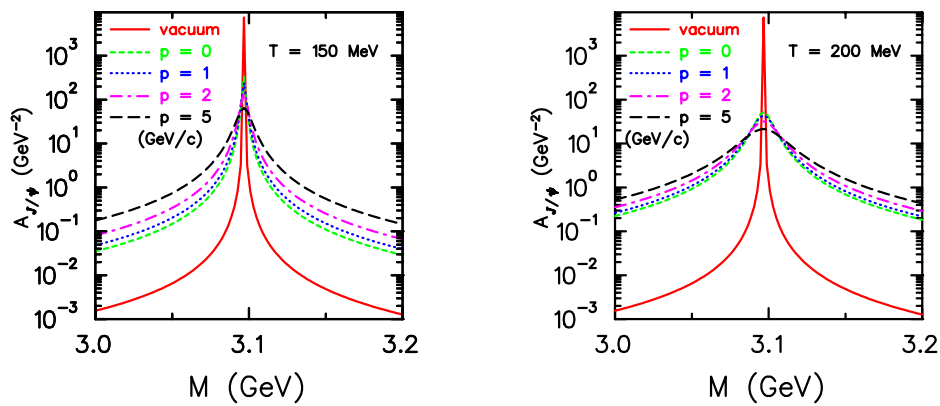,width=4.5cm}
\hspace{0.05cm}
\end{center}
\caption{Spectral function in vacuum and at finite temperature
             for fixed momenta and fixed temperatures.}
\label{sfunction}
\end{figure}

The spectral function resulting from this calculation
appears in Fig.~\ref{sfunction}.  We fix the temperature at
values of 150 and 200 MeV, which are relevant values for heavy 
ion experiments.
Four different values of $J/\psi$ momenta are chosen for each.
We analyze the spectral function this way since there is no expectation
that the $J/\psi$'s will be thermal.
The spectral function for the $J/\psi$ is clearly modified
at finite temperature due to interactions with hadronic
matter.  Breakup rates fed into the spectral function
range from 5 to 30 MeV.

\section{Conclusions}

We have applied an effective Lagrangian to describe the dynamics of
$J/\psi$ with particular focus on breakup reactions
induced by light unflavoured mesons.   Off-shell
effects at the vertices were handled with QCD sum rule estimates for
the running coupling constants, {\it\/i.e.\/}
the three-point functions.  Analytic continuation to the poles (on-shell
mesons) gave further guidance to constrain couplings. 
Four-point functions were first expanded over general Lorentz
structures and then constrained with Ward identities.
The formalism is then covariant, gauge invariant, and consistently
describes on a graph-by-graph basis the kinematics and the
off-shell vertex functions.  The model uses effective field theory,
QCD sum rules for the form factors, and Ward identities to constrain
the overall scattering amplitudes.

Resulting cross sections for breakup of $J/\psi$ are several millibarns,
and exhibit nontrivial energy dependence.  When imported to
the spectral function, the resulting effects give a dramatic
modification at finite temperature as compared to vacuum 
structure.  This could have important consequences for studies at
SPS and RHIC measuring dilepton signals near the $J/\psi$ mass and 
impact interpretation of $J/\psi$ suppression.

\section*{Acknowledgements}
This work has been supported in part by the Natural Sciences and Engineering
Research Council of Canada, in part by the Fonds Nature et Technologies of
Quebec and by the National Science Foundation under grant number 
PHY-0098760.

\end{document}